\documentclass[onecolumn]{aa} 

\usepackage{multirow}
\usepackage{graphicx}
\usepackage{txfonts}
\usepackage{lscape}
\begin{document}

   \title{A spectral line survey of IRC +10216 between 13.3 and 18.5 GHz}


   \author{Xiao-Yan Zhang\inst{1,2}
          \and
          Qing-Feng Zhu\inst{1,2}
          \and
          Juan Li\inst{3,4}
          \and
          Xi Chen\inst{5,3,4}
          \and
          Jun-Zhi Wang\inst{3,4}
          \and
          Jiang-Shui Zhang\inst{5}
          }

\institute{CAS Key Laboratory for Research in Galaxies and Cosmology, Department of Astronomy, University of Science and Technology of China, Hefei 230026, China\\
  	     \email{xyz1128@mail.ustc.edu.cn, zhuqf@ustc.edu.cn}
   	\and
School of Astronomy and Space Science, University of Science and Technology of China, Hefei 230026, China
	\and
	Department of Radio Science and Technology, 
   	Shanghai Astronomical observatory, 80 Nandan RD, Shanghai 200030, China\\
              \email{lijuan@shao.ac.cn}
         \and
            Key Laboratory of Radio Astronomy, Chinese Academy of Sciences, China
            \and
            Center for Astrophysics, Guangzhou University, Guangzhou 510006, China
             }

   \date{}

\abstract {
	A spectral line survey of IRC +10216 between 13.3 and 18.5 GHz is carried out 
	              using the Shanghai Tian Ma 65 m Radio Telescope (TMRT-65m) with a sensitivity of < 7 mK. 
	Thirty-five spectral lines of 12 different molecules and radicals are detected in total. 
	Except for SiS, the detected molecules are all carbon-chain molecules, 
	             including HC$_3$N, HC$_5$N, HC$_7$N, HC$_9$N, C$_6$H, C$_6$H$^-$, C$_8$H, SiC$_2$, SiC$_4$, c-C$_3$H$_2$ and l-C$_5$H. 
	The presence of rich carbon-bearing molecules is consistent with the identity of IRC +10216 as a carbon-rich AGB star.
	The excitation temperatures and column densities of the observed species are derived by assuming a local thermodynamic equilibrium and homogeneous conditions.
 	}

   \maketitle
%

\section{Introduction}
IRC+10216 (CW Leo) is a well-known carbon-rich Asymptotic Giant Branch (AGB) star 
	with a high mass-loss rate of about $1–-4\times10^{-5}\ \mathrm{M_{\sun} \ yr^{-1}}$ (e.g., \citealt{crosas1997physical,Beck2012On,cernicharo2015molecular})
	at a distance of $123 \pm 14\,\mathrm{pc}$ \citep{groenewegen2012independent}. 
The effective stellar temperature is between 2500 K and 2800 K \citep{men2001structure}.
The radial velocity (LSR) and terminal expansion velocity of IRC +10216 have been estimated 
	in $-26.5\pm0.3\ \mathrm{km\ s^{-1}}$ and $14.5\pm0.2\ \mathrm{km\ s^{-1}}$, respectively
	\citep{cernicharo2000lambda}. 
IRC+10216 is surrounded by an extended circumstellar envelope (CSE) owing to its extensive mass loss. 
The CSE is a physically and chemically rich environment to form molecules and condense dust grains (e.g., \citealt{fonfria2008detailed,agundez2012molecular,fonfria2014complex}). 

Observations at centimeter (cm) and millimeter (mm) wavelengths probe the cool, 
	outer part of AGB winds and bring wealthy evidence of molecular species, 
	from simple radicals such as OH to more complex species like cyanopolyynes. 
In previous works, linear carbon-chain molecules like C$_n$H, HC$_{2n+1}$N, and C$_n$S are observed to be abundant in the circumstellar envelopes of carbon-rich AGB stars \citep{winnewisser1978detection, cernicharo1996discovery, guelin1997detection,Ag2014New,gong20151,Ag2017Growth}. 
These molecules are proposed to be related to the formation and destruction of polyaromatic hydrocarbons (PAH) 
	\citep{henning1998carbon, tielens2008interstellar}. 
The full characterization of carbon-chain molecules is regarded to be an important issue in astrochemistry \citep{sakai2010abundance}.

Molecular line survey is a powerful tool for analyzing both physical and chemical parameters of astronomical objects. 
Several systematical spectral line surveys of IRC+10216 have been reported in the literature. 
From Table \ref{Surveys}, we find that existing surveys cover the frequency range from 4 GHz to 636.5 GHz for IRC +10216. 
Over 80 species \citep{Ag2014New} have been discovered toward IRC +10216, 
	including unusual carbon-chain and silicon-carbon molecules such as SiC$_2$, MgCCH, NCCP and SiCSi
	\citep{Thaddeus1984Identification,cernicharo2010high,
	Ag2014New,Cernicharo2015Discovery}, 
metal cyanides/isocyanide such as MgNC, MgCN, AlNC, KCN, FeCN, NaCN, HMgNC and SiH$_3$CN
	\citep{Ziurys1995DETECTION,ziurys2002more,Pulliam2010Identification,
	Zack2011Detection,agundez2012molecular,Cabezas2013Laboratory,Ag2014New}, 
	and even metal halides such as NaCl, KCl, AlCl and AlF \citep{cernicharo1987metals,Ziurys1995DETECTION,Quintanalacaci2016Hints}.

Complex molecules have small rotational constants so that their lowest-energy transitions arise at cm wavelengths. 
From Table \ref{Surveys}, we find that there are several systematic and high sensitivity surveys at cm wavelengths toward IRC +10216. 
However, no systematic survey has been reported in the frequency range between 6 and 17.8 GHz.

In this paper, the results of a spectral line survey of IRC +10216 between 13.3 and 18.5 GHz are presented. 
The observations are introduced in Section 2. 
The observational results derived from the analysis of the molecular emission are shown in Section 3. 
Section 4 contains the analysis of the lines, the comparison of the results with those of previous works. 
Finally, the conclusions are presented in Section 5.

\newcommand{\tabincell}[2]{\begin{tabular}{@{}#1@{}}#2\end{tabular}}
\renewcommand{\multirowsetup}{\centering}

\begin{table*}   
\caption{\label{Surveys}Existing line surveys of IRC +10216.}
\centering
\begin{tabular}{cc ccc}
\hline\hline
\tabincell{c}{Frequencies \\(GHz)} & Telescope & Reference & \tabincell{c}{Sensitivity (1$\sigma$)\\(mJy)} 
	& \tabincell{c}{FWHM beam or\\Angular Resolution}($\arcsec$)\tablefootmark{$\triangle$}\\
\hline
4--6 & Arecibo-305m & \citet{araya2003c} & 0.3 & 46—-68\\ 
\hline
\bfseries 13.3--18.5 & \bfseries TMRT-65m & \bfseries This work & \bfseries 12 & \bfseries 52-—73 \\
\hline
17.8--26.3 & Effelsberg-100m & \citet{gong20151} & 1.0 &  35-—50 \\ 
\hline
28--50 & Nobeyama-45m & \citet{kawaguchi1995spectral} & 35 &  35—-62 \\ 
\hline
72--91 & Onsala-20m &  \citet{johansson1984spectral,johansson1985spectra} & 665 &  41—-52 \\
\hline
80–-116 & \multirow{3} {*} {IRAM-30m}\tablefootmark{$\dagger$} & \citet{Ag2014New}  & 23  & 21-—31   \\ 
\cline{3-5}
129.0—-172.5 &   & \citet{cernicharo2000lambda} & 78  & 14--19 \\ 
\cline{3-5}
197--357.5 &  & \citet{Ag2008Tentative} &  39   &  7--13  \\ 
\hline
84.0--115.5 & ALMA & \citet{Ag2017Growth} &  1.3  & 1.0 \\  
\hline
131.2--160.3 & ARO-12m & \multirow{2} {*} {\citet{he2008spectral}} & \multirow{2} {*} {1400} &  39—-47  \\  
219.5--267.5\tablefootmark{$\ast$} & SMT-10m &  &  &  28—-34 \\
\hline
214.5–-285.5 & SMT-10m & \citet{Tenenbaum2010The} & 140   &  26—-35 \\
\hline
253--261 & CARMA & \citet{fonfria2014complex}  & 20 &  0.25 \\ 
\hline
\multirow{2} {*} {255.3–-274.8} & \multirow{2} {*} {ALMA} & \citet{Cernicharo2013Unveiling} & 5 & 0.6 \\  
  &  & \citet{Prieto2015SI} & 17 & 0.6  \\ 
\hline
293.9--354.8 & SMA & \citet{patel2011interferometric} & 33 & 3 \\ 
\hline
330.2–-358.1 & CSO-10.4m & \citet{groesbeck1994molecular} & 6000  & 20—-22 \\ 
\hline
222.4--267.9\tablefootmark{$\diamond$} & \multirow{2} {*} {JCMT-15m} & \multirow{2} {*} {\citet{avery1992spectral}} &   \multirow{2} {*} {3100} &  18—-22 \\  
339.6--364.6 &    &    &   &  15—-16  \\ 
\hline
554.5--636.5 & Herschel/HIFI & \citet{cernicharo2010high} &  3400  &  33—-38  \\ 
\hline
\end{tabular}
\tablefoot{
\tablefoottext{$\ast$}{The frequency range from 219.5 GHz to 267.5 GHz was discontinuously covered.}\\
\tablefoottext{$\diamond$}{The frequency range from 222.4 GHz to 267.9 GHz was discontinuously covered.}\\
\tablefoottext{$\triangle$}{The FWHM beam sizes are listed for single-dish, the angular resolutions are listed for interferometers.}\\
\tablefoottext{$\dagger$}{There are some other surveys using of IRC +10216 IRAM 30m, although only partially published,
e.g., \citet{cernicharo1987metals,cernicharo1996discovery,guelin1997detection,Gu2000Astronomical,ziurys2002more,Gu2004Detection,Cernicharo2007Astronomical,Ag2007Discovery,Cernicharo2008Detection,Agundez2008Detection,cernicharo2011probing,agundez2012molecular,Cernicharo2015Discovery}.}
}
\end{table*}

\section{Observations and data reduction}
The Tian Ma Radio Telescope (TMRT) is a 65m diameter fully-steerable radio telescope located in the western suburbs of Shanghai, China \citep{li2016tmrt, yan2015single}. 
The Digital Backend System (DIBAS) of TMRT is an Field Programmable Gate Array (FPGA) based spectrometer 
	based upon the design of Versatile GBT Astronomical Spectrometer (VEGAS).

The observations were performed in a position-switching mode at Ku band (11.5--18.5 GHz) 
	towards IRC +10216 in 2016 March and April with the TMRT. 
On-source and off-source integration times were two minutes per scan. 
In this work, the DIBAS sub-band mode 2 with a single spectral window was adopted for Ku band observation. 
The window has 16384 channels and a bandwidth of 1500 MHz, 
	supplying a velocity resolution of about $2.08\  \mathrm{km\ s^{-1}}$ (13.3 GHz) 
	to $1.491\ \mathrm{km\ s^{-1}}$ (18.5 GHz). 
The intensity was calibrated by injecting periodic noise, 
	and the accuracy of the calibration was estimated from frequency 
	ranges where the spectrum was apparently free of lines. 
The system temperature was about 55 -- 64 K at Ku band. 
Across the whole frequency range, the FWHM beam size was 52\arcsec -- 84\arcsec. 
The adopted coordinates for our searches were: R.A. (2000) = 09:47:57.45, DEC (2000) = 13:16:43.8. 
The pointing accuracy was better than 12\arcsec. 
The resulting antenna temperatures were scaled to main beam temperatures ($T_{MB}$) 
	by using a main beam efficiency of 0.6 for a moving position of sub-reflector at Ku band \citep{li2016tmrt}.

The data were reduced using GILDAS software package\footnotemark[1]\footnotetext[1]{http://www.iram.fr/IRAMFR/GILDAS.} 
	including CLASS and GREG. 
Linear baseline subtractions were used for all the spectra. 
Because of the contamination due to time variable radio frequency interference (RFI), 
	the channels from 11.5 GHz to 13.3 GHz were discarded from further analysis. 
Since some unknown defects occurred at the edges of the spectra, 
	the channels with a bandwidth of 150 MHz at each edge were excluded. 
All the spectra including two polarizations were averaged to reduce rms noise levels.


\section{Results}
 \begin{figure*}
   \centering
  \includegraphics[width=\hsize]{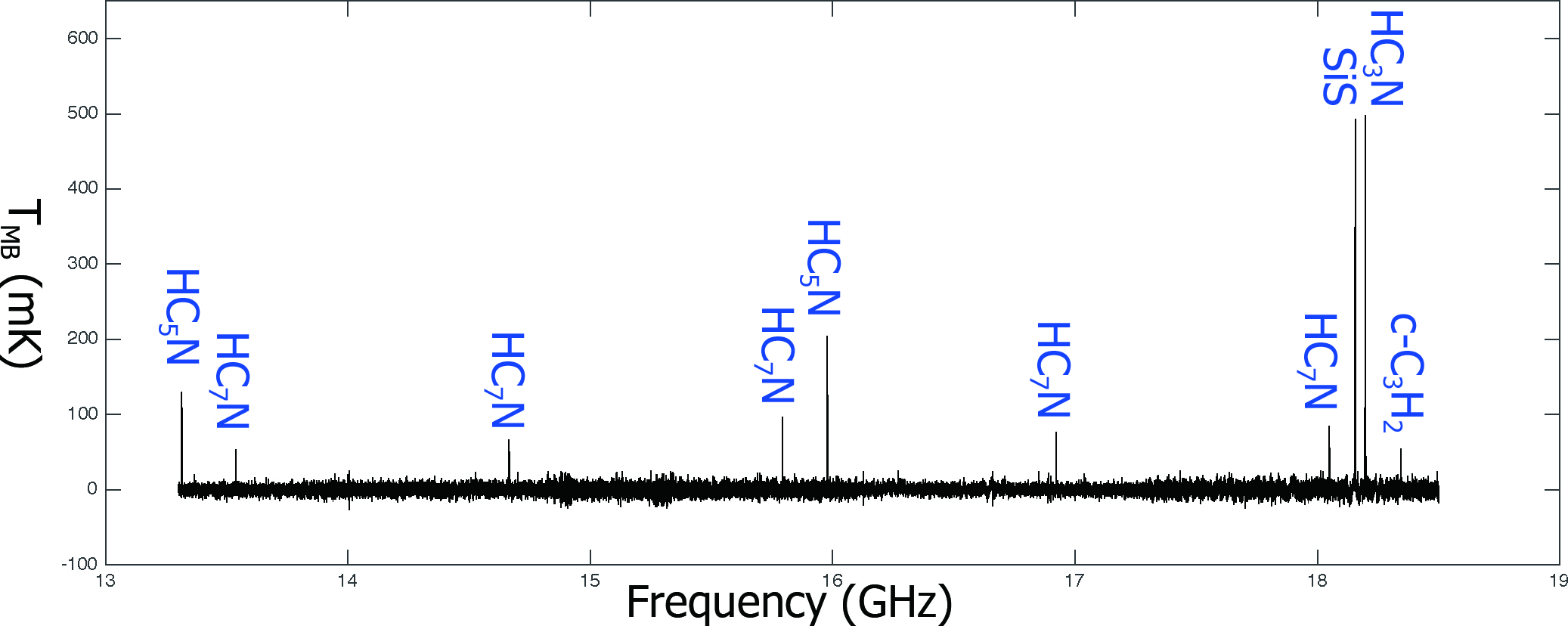}
   \caption{ Overview of the spectral line survey of IRC +10216 between 13.3 and 18.5 GHz with strong lines marked.
  }
   \label{Overview}%
    \end{figure*}

 \begin{figure}
   \centering
   \resizebox{\hsize}{!}{\includegraphics{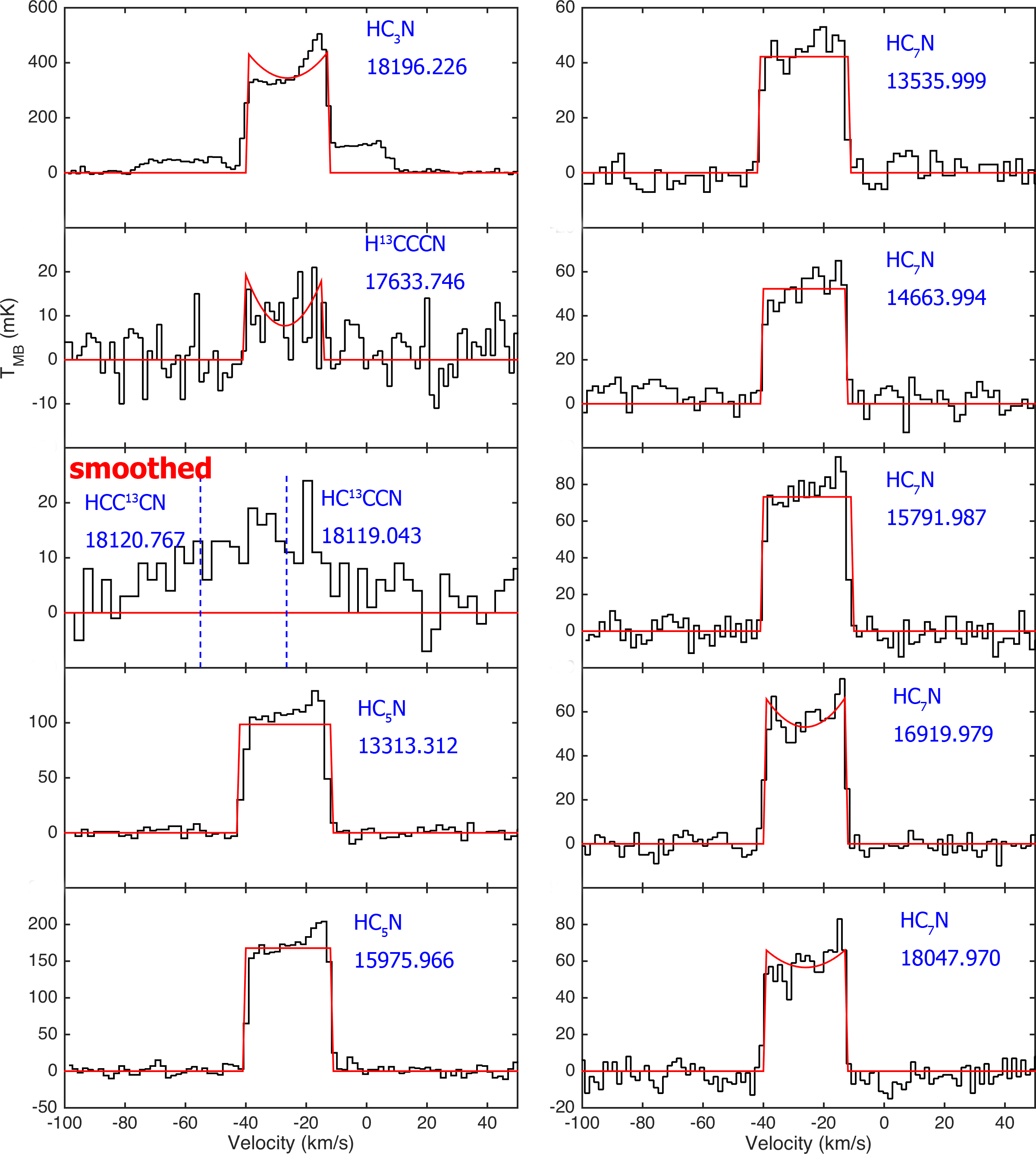}}
   \caption{(a) Zoom in of all detected and tentatively detected lines for HC$_3$H, H$^{13}$CCCN, HC$^{13}$CCN, HCC$^{13}$CN, HC$_5$H and HC$_7$H. 
   The corresponding rest frequency in MHz are shown in the upper right of each panel. 
   Weak lines have been smoothed to have a channel width of 3.0--3.6 $\mathrm{km\ s^{-1}}$, 
   	and are marked with ``smoothed'' in the upper left of the corresponding panels. 
   Otherwise, the channel width is 1.5--2.1 $\mathrm{km\ s^{-1}}$.
The blue dashed lines of the blended transitions trace the systematic LSR velocity ($-26.5\ \mathrm{km\ s^{-1}}$) of IRC +10216. }
    \label{Fig1a}
    \end{figure}

    \addtocounter{figure}{-1}
\begin{figure}
   \centering
\resizebox{\hsize}{!}{\includegraphics{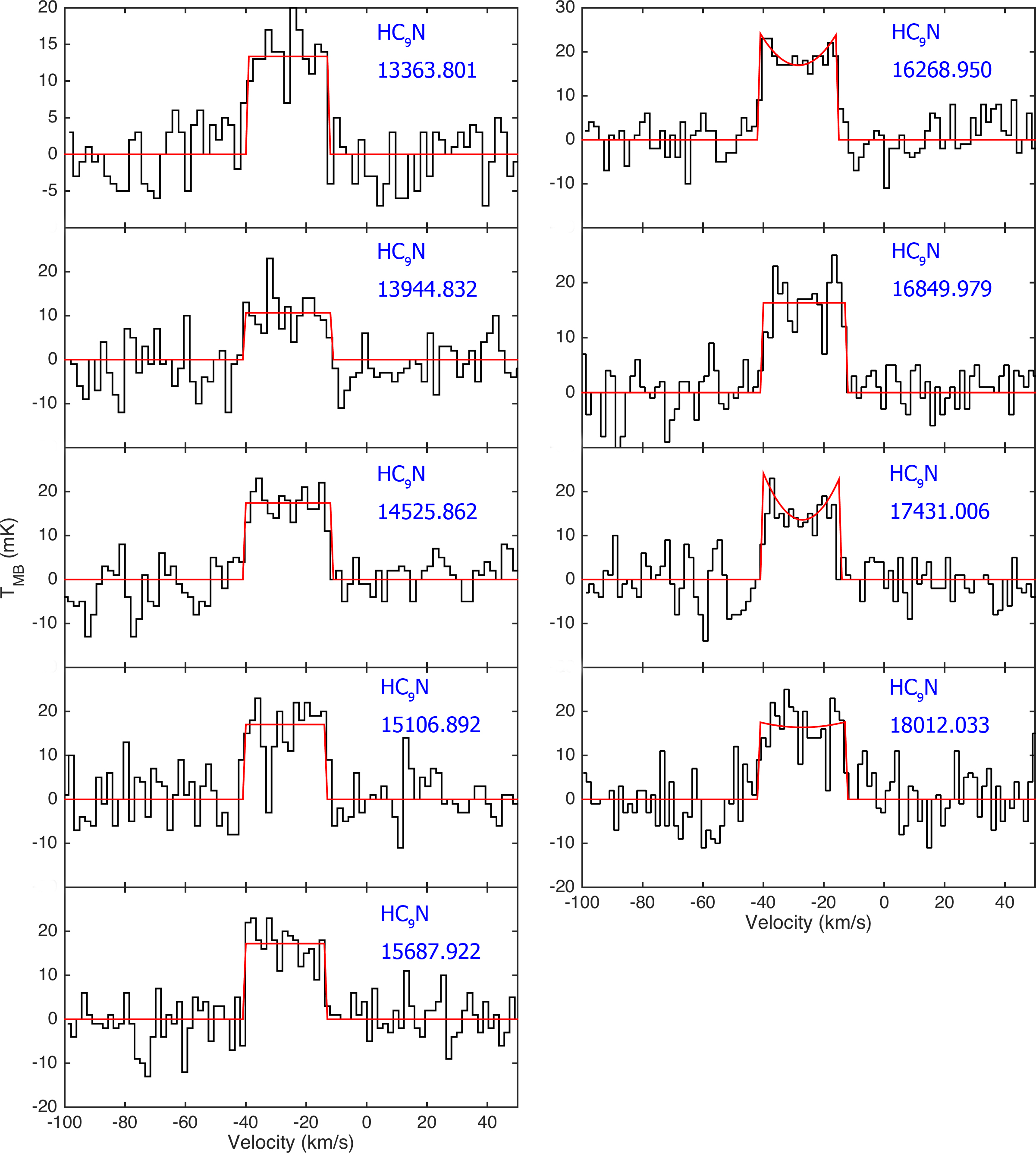}}
\caption{(b) Same as Fig.2.(a) for HC$_9$H.}
\label{Fig1b}
 \end{figure}

 \addtocounter{figure}{-1}

 \begin{figure}
   \centering
 \resizebox{\hsize}{!}{\includegraphics{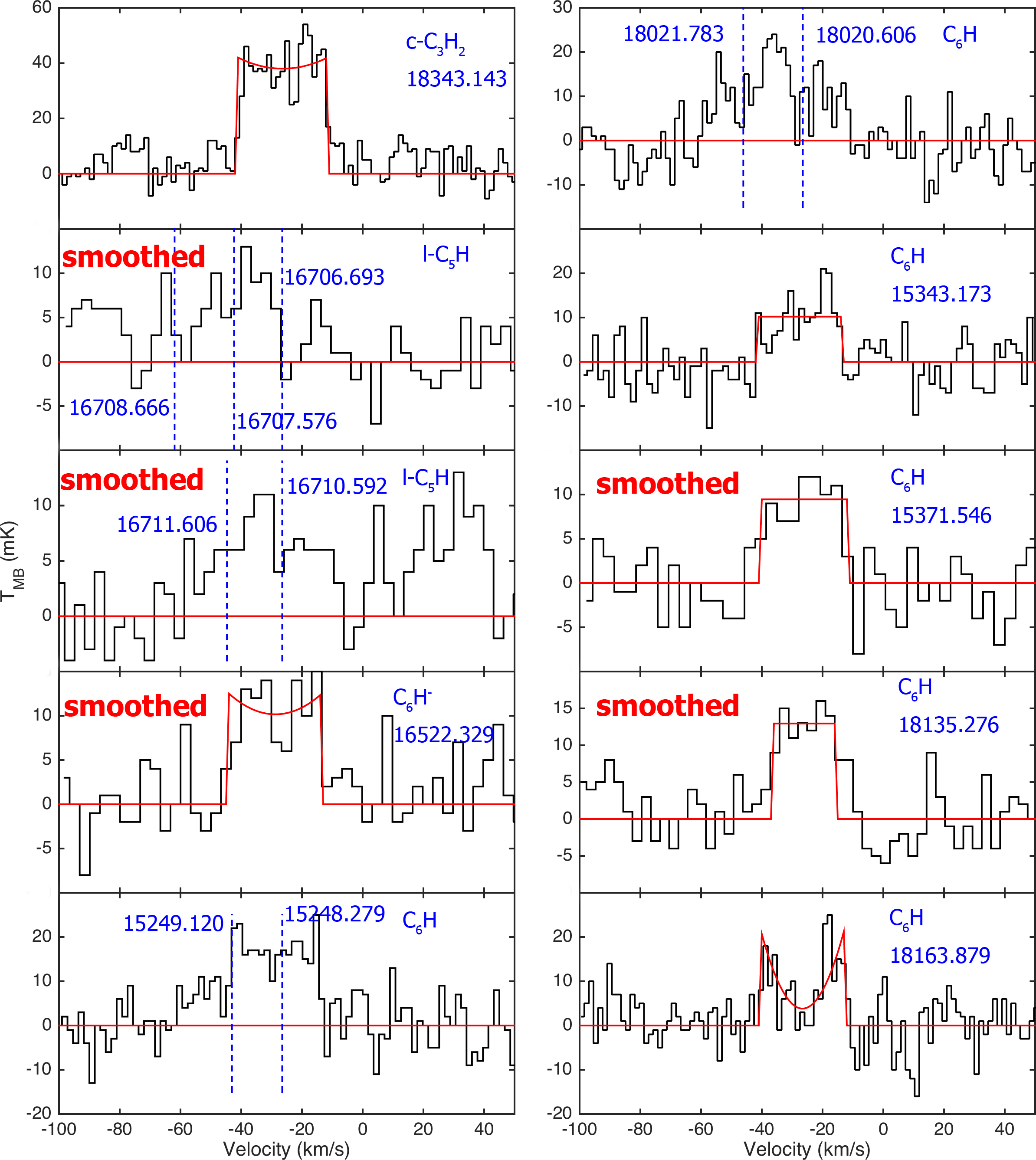}}
   \caption{(c) Same as Fig.2.(a) for C$_n$H}
\label{Fig1c}
 \end{figure}

 \addtocounter{figure}{-1}

    \begin{figure}
   \centering
 \resizebox{\hsize}{!}{\includegraphics{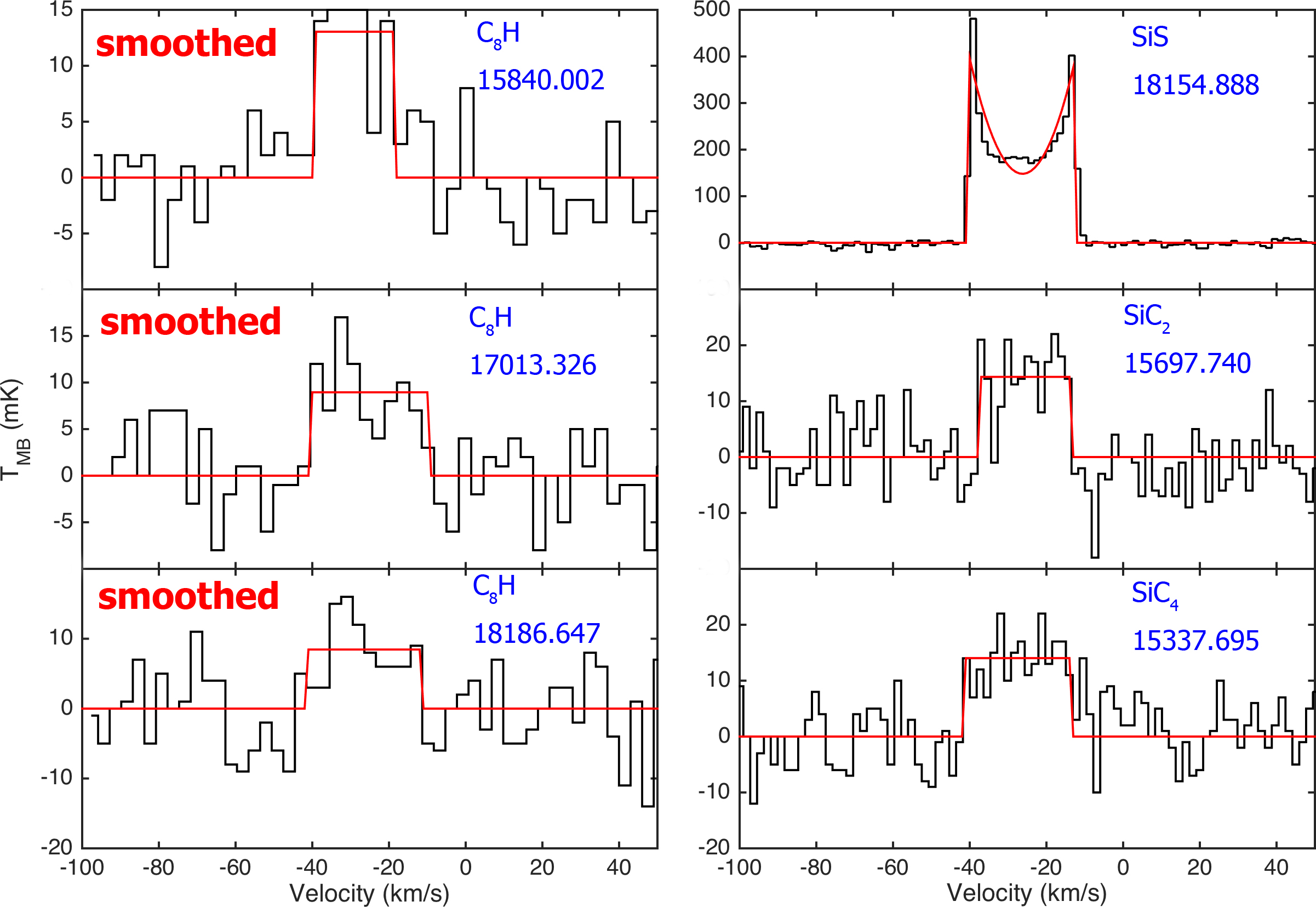}}
   \caption{(d) Same as Fig.2.(a) for C$_8$H and Si bearing species.}
\label{Fig1d}
 \end{figure}
Based on the molecular database ``Splatalogue''\footnotemark[2]\footnotetext[2]{www.splatalogue.net.}, 
	which is a compilation of the Jet Propulsion Laboratory (JPL, \citealt{pickett1998submillimeter}), 
	Cologne Database for Molecular Spectroscopy catalogues (CDMS, \citealt{muller2005cologne}), 
	and Lovas/NIST catalogues \citep{lovas2004nist}, 
	the line identifications are performed. 
The local standard of rest (LSR) radial velocity of $-26.5\ \mathrm{km\ s^{-1}}$ is adopted to derive the rest frequency of the observed lines. 
A line is considered real if it has a signal-to-noise ratio (S/N) of at least five \citep{gong20151}, 
although, a line is already significant if it is characterized by a S/N above three \citep{fonfria2014complex}.
Lines with a S/N of at least three are discussed in this paper.
Figure \ref{Overview} presents an overview of the spectral line survey of IRC +10216 between 13.3 and 18.5 GHz with strong lines marked.

All detected transitions are shown in Figure \ref{Fig1a} and Table \ref{Parameters}.
There are 41 transitions assigned to 12 different molecules and radicals found in this survey. 
Except for SiS, all other molecules are C-bearing molecules. 
The detected transitions include one transition of HC$_3$N, 
	three transitions of the $^{13}$C-bearing isotopologues of HC$_3$N, 
	two transitions of HC$_5$N, five transitions of HC$_7$N, 
	and nine transitions of HC$_9$N, 
	which are shown in Figure \ref{Fig1a}(a) and Figure \ref{Fig1b}(b).
In Figure \ref{Fig1a}(a), HC$^{13}$CCN ($v=0, J=2-1$) is blended with HCC$^{13}$CN ($v=0,\ J=2-1$). 
Additionally, eight transitions of C$_6$H are detected. 
In Figure \ref{Fig1c}(c), C$_6$H ($ ^2 \Pi_{3/2}, \  J=11/2-9/2,\ l=f$) is blended with C$_6$H ($^2\Pi_{3/2},\  J=11/2-9/2,\ l=e$), 
	and C$_6$H ($^2\Pi_{3/2},\ J=13/2-11/2,\ l=f$) is blended with C$_6$H ($^2\Pi_{3/2},\ J=13/2-11/2,\ l=e$). 
There is a transition of c-C$_3$H$_2$,
	five transitions of l-C$_5$H from the $^2\Pi_{1/2}$ ladder 
	and one transition of C$_6$H$^-$ in Figure \ref{Fig1c}(c).
Spectra line profiles of three transitions of C$_8$H from the $^2\Pi_{3/2}$ ladder, 
	SiS ($v=0, J=1-0$), SiC$_2$ ($v=0, J=7(2,5)-7(2,6)$) and SiC$_4$ ($v=0, J=5-4$)
	are shown in Figure \ref{Fig1d}(d).
There are five transitions of SiC$_2$ and two transitions of SiC$_4$ in this band. 
But other transitions of SiC$_2$ and SiC$_4$ are too weak to be detected.
The peak intensities of the smoothed lines of HC$^{13}$CCN, HCC$^{13}$CN, l-C$_5$H, 
	C$_8$H $^2\Pi_{3/2}$ and C$_6$H $^2\Pi_{1/2}$ are smaller than 20 mK.
	These smoothed lines are weak and tentatively detected.
The rest frequencies and the upper energy levels  of these lines 
	are obtained from the molecular database ``Splatalogue''.

The SHELL fitting routine in Continuum and Line Analysis Single-dish Software (CLASS) 
	is used to derive line parameters including peak intensity, 
       integrated intensity, and expansion velocity which is defined as the half-width at zero power. 
Except for the blended lines, the observed lines are either double peaked or flat-topped.
The fitting profiles are shown in Figure \ref{Fig1a} when possible. 
For the line of HC$_3$N with hyperfine structure, only the main component is fitted to obtain the expansion velocity. 
For the lines that are blended and weak, the parameters are estimated directly by integrating the line profiles. 
Firstly, estimate the maximum velocities $\upsilon_1$ and $\upsilon_2$ of blueshift and redshift.
Then, return the integrated area ($\mathrm{K\ km/s}$) of the current spectrum 
	between velocities $\upsilon_1$ and $\upsilon_2$ by using the function TDV($\upsilon_1,\upsilon_2$) in CLASS.
The observed properties of the lines are displayed in Table \ref{Parameters}.

\longtab{
\begin{landscape}
\begin{longtable}{ccccc ccccc c}
\caption{\label{Parameters}Line Parameters of Transitions Detected in IRC +10216.}\\
\hline\hline
(1)  & (2) &(3)&(4)&(5)&(6)&(7)&(8)&(9)&(10)&(11)\\
Species & Transitions &   \tabincell{c}{Rest Freq.\\ (MHz)} &  \tabincell{c}{$E_U/k$\\ (K)} & Observing date &   \tabincell{c}{FWHM \\ beam (\arcsec)}  &\tabincell{c}{peak intensity \\(mK)}   & \tabincell{c}{$v_{exp}$\\ $(\mathrm{km\ s^{-1}})$}  & \tabincell{c}{LSR velocity\\ ($\mathrm{km\ s^{-1}}$) } &\tabincell{c}{$\int {T_{MB}} \,\mathrm{d}\upsilon$\\ $(\mathrm{mK\ km\ s^{-1}})$}  & Notes\\
\hline
\endfirsthead
\caption{continued.}\\
\hline\hline
(1)  & (2) &(3)&(4)&(5)&(6)&(7)&(8)&(9)&(10)&(11)\\
Species & Transitions &   \tabincell{c}{Rest Freq.\\ (MHz)} &  \tabincell{c}{$E_U/k$\\ (K)} & Observing date &   \tabincell{c}{FWHM \\ beam (\arcsec)}  &\tabincell{c}{peak intensity \\(mK)}  & \tabincell{c}{$v_{exp}$\\ $(\mathrm{km\ s^{-1}})$}  & \tabincell{c}{LSR velocity\\ ($\mathrm{km\ s^{-1}}$) } &\tabincell{c}{$\int {T_{MB}} \,\mathrm{d}\upsilon$\\ $(\mathrm{mK\ km\ s^{-1}})$}& Notes \\
\hline
\endhead
\hline
\endfoot
HC$_3$N &  $v=0,\ J=2-1$  & 18196.2260 & 1.30994 & 2016Apr10 & 53 &504.3(6.8)  & 13.7(1.5)&-26.2(1.5)& 10337(197) & D \\
H$^{13}$CCCN& $v=0,\ J=2-1$ & 17633.7460 &1.26928 &2016Apr10 & 55 &27.2(5.4)  & 13.1(1.6)& -27.1(1.6) & 306(157) & D  \\
HC$^{13}$CCN& $v=0,\ J=2-1$ & 18119.0430 &1.30437 &2016Apr10 & 54 &\multirow{2}{*}{24.3(3.3)}&  & &\multirow{2}{*}{711(96)} &\multirow{2}{*}{B,S}  \\
HCC$^{13}$CN& $v=0,\ J=2-1$ & 18120.7670 &1.30445 &2016Apr10 & 54 & & & & &  \\
\cline{2-11}
\multirow{2}{*}{HC$_5$N} &  $v=0,\ J=5-4$  & 13313.3119 & 1.91685 & 2016Apr10 & 73 & 128.9(3.5)& 15.5(2.1)&-27.4(2.1)&3063(102) & F \\
  & $v=0,\ J=6-5$ & 15975.9663 & 2.68359  & 2016Mar25 & 61 & 203.7(5.5) & 14.5(1.7)&-26.2(1.7)&4870(160)& F \\
\cline{2-11}
 \multirow{5}{*}{HC$_7$N} & $v=0,\ J=12-11$ & 13535.9990 &  4.22252 & 2016Apr10 & 72 & 52.6(3.9)  &15.2(2.0)&-26.3(2.0)&1279(113)& F \\
  & $ v=0,\ J=13-12$ & 14663.9936 &  4.92625 & 2016Mar25 & 66 & 64.7(5.4)  & 14.0(1.9)&-26.4(1.9)&1463(157)& F \\
  & $v=0,\ J=14-13$ & 15791.9870 &  5.68424 & 2016Mar25 & 61 &94.8(5.7) & 14.8(1.7)&-25.6(1.7)& 2163(165)& F \\
  & $v=0,\ J=15-14$ & 16919.9791 &  6.49617 & 2016Mar10 & 57 & 74.5(4.4)  &13.6(1.6)&-26.1(1.6)& 1578(128)& D \\
  & $v=0,\ J=16-15$ & 18047.9697 &  7.36235 & 2016Apr10 & 54 & 82.6(6.7)   & 13.6(1.5)&-26.0(1.5)&1629(194)& D \\
\cline{2-11}
 \multirow{9}{*}{HC$_9$N} & $v=0,\ J=23-22$ & 13363.8008 & 7.69636 & 2016Apr10 & 73& 20.0(3.9)   & 13.4(2.1)&-26.3(2.1)& 357(113)& F \\
  & $v=0,\ J=24-23$ & 13944.8317 & 8.36550 & 2016Mar25 & 70 & 22.9(5.4)  & 14.8(2.0)&-26.2(2.0)& 314(157)&F \\
  & $v=0,\ J=25-24$  & 14525.8622 & 9.06270 & 2016Mar25 & 67 & 23.4(4.9)  & 14.3(1.9)&-26.2(1.9) & 496(142) & F  \\
  & $v=0,\ J=26-25$ & 15106.8921 & 9.78767 & 2016Mar25 & 64 & 23.0(5.7)   & 13.7(1.8)& -26.9(1.8)& 468(165)& F \\
  & $v=0,\ J=27-26$ & 15687.9215 & 10.54055 & 2016Mar25 & 62 & 28.6(5.1) &13.1(1.8)&-27.1(1.8)&451(148)& F  \\
  & $v=0,\ J=28-27$ & 16268.9503 & 11.32135 & 2016Mar10 & 60 & 23.4(3.9) & 12.8(1.7)&-28.4(1.7)&498(113)& D \\
  & $v=0,\ J=29-28$ & 16849.9786 & 12.13005 & 2016Mar10 & 58 & 24.6(3.8) &13.9(1.6)&-26.3(1.6)&453(110)& F  \\
  & $v=0,\ J=30-29$ & 17431.0062 & 12.96652 & 2016Apr10 & 56 & 33.5(5.8)  & 13.0(1.6)&-27.0(1.6)&447(168)& D \\
  & $v=0,\ J=31-30$ & 18012.0333 & 13.83105 & 2016Apr10 & 54 & 24.7(5.5) &14.4(1.5)&-27.2(1.5)&483(160)& D\\
  \hline
c-C$_3$H$_2$&$v=0,\ J=1(1,0)-1(0,1)$&18343.1430&3.23012&2016Apr10&53&53.8(6.1) &14.8(1.5)&-26.3(1.5)&1161(157)& D \\
\cline{2-11}
\multirow{5}{*}{l-C$_5$H $^2\Pi_{1/2}$}&$J=7/2-5/2,F=4-3,l=e$&16706.6928&1.71872&2016Mar16&58 &\multirow{3}{*}{13.4(3.1)}& & &\multirow{3}{*}{314(90)} & \multirow{3}{*}{B,S} \\
  &$J=7/2-5/2,F=3-2, l=e$&16707.5764&1.71876&2016Mar16&58 & &   & & & \\
  &$J=7/2-5/2,F=3-3, l=e$&16708.6662&1.71881&2016Mar16&58 & &   & & & \\
  &$J=7/2-5/2,F=4-3, l=f$&16710.5921&1.71962&2016Mar16&58 &\multirow{2}{*}{11.1(3.1)}& & & \multirow{2}{*}{313(90)}& \multirow{2}{*}{B,S}\\
  &$J=7/2-5/2,F=3-2, l=f$&16711.6058&1.71924&2016Mar16&58  & & &   &  \\
\cline{2-11}
C$_6$H$^-$ &$J=6-5$&16522.3290&2.77528&2016Mar16&59 &14.6(4.0)  &15.7(3.3)&-28.7(3.3)&345(116)& D,S \\
\cline{2-11}
\multirow{4}{*}{C$_6$H $^2\Pi_{1/2}$} & $J=11/2-9/2,\ l=f$&15343.1726&24.51330&2016Mar25&63&20.9(5.3) &14.3(1.8)&-27.7(1.8)&291(154)& F\\
 &$J=11/2-9/2,\ l=e$&15371.5462&24.52142&2016Mar25&63 &12.1(3.8) &14.4(3.6)&-26.0(3.6)& 272(110) & F,S \\
  &$J=13/2-11/2,\ l=f$&18135.2758&25.38366&2016Apr10&54&16.1(3.6) &10.6(3.0)&-26.4(3.0)& 275(104) &F,S \\
  &$J=13/2-11/2,\ l=e$&18163.8788&25.39309&2016Apr10&53&25.3(5.4) &13.7(1.5)&-26.6(1.5)&264(157)& D\\
\cline{2-11}
  \multirow{4}{*}{C$_6$H $^2\Pi_{3/2}$} & $J=11/2-9/2,\ l=e$ &15248.2791&2.12884&2016Mar25&64 &\multirow{2}{*}{33.2(5.3)}& & &\multirow{2}{*}{704(154)}& \multirow{2}{*}{B}\\
  & $J=11/2-9/2,\ l=f$&15249.1203&2.12903&2016Mar25&64& & & & & \\
  & $J=13/2-11/2,\ l=e$&18020.6059&2.99379&2016Apr10&54&\multirow{2}{*}{24.4(6.7)}& & &\multirow{2}{*}{576(194)} & \multirow{2}{*}{B} \\
  & $J=13/2-11/2,\ l=f$&18021.7828&2.99385&2016Apr10&54 & & & &  & \\
\cline{2-11}
\multirow{3}{*}{C$_8$H $^2\Pi_{3/2}$}&$J=27/2-25/2$&15840.0020&5.40584&2016Mar25&61&14.8(3.7) & 10.6(3.5)& -29.1(3.5) & 277(107)&F,S\\
 &$J=29/2-27/2$&17013.3256&6.22239&2016Mar16&57&17.1(3.6) &15.4(3.2)&-25.0(3.2)& 276(104)&F,S\\
 &$J=31/2-29/2$&18186.6470&7.09520&2016Mar16&53&15.8(5.1) &15.0(3.0)&-26.4(3.0)& 254(148)&F,S\\
\hline
SiS&$v=0,\ J=1-0$&18154.8880&0.87129&2016Apr10&53&480.6(6.4)  &14.2(1.5)&-26.2(1.5)&6692(186)& D \\
SiC$_2$&$v=0,\ J=7(2,5)-7(2,6)$&15697.7396&40.16052&2016Mar25&62 &22.3(5.6) &12.2(1.7)&-25.6(1.7)&350(162)& F\\
SiC$_4$&$J=5-4$&15337.6950&2.20824&2016Mar25&63&22.3(5.3) &14.3(1.8)&-27.4(1.8)&400(154)& F\\
\end{longtable}
\tablefoot{
The numbers in parentheses represent the $1—\sigma$ errors. 
The rms noise of peak intensity is estimated from frequency ranges 
	around each particular transition where the spectrum is apparently free of lines. 
The error for expansion velocity and LSR velocity is the velocity width of a channel. 
And the error of integrated intensity given here is 
	the product of the rms noise of peak intensity and the width of the line for about $29\ \mathrm{km\ s^{-1}}$. 

Each column represents the following information, Col. (1): molecule name; Col. (2): transition; Col. (3): rest frequency; Col. (4): $E_U/k$, the upper level energy in K; Col. (5): observing date; Col. (6): FWHM beam; Col. (7): peak intensity of main beam temperature; Col. (8): expansion velocity; Col. (9): LSR velocity; Col. (10): integrated intensity.
(D) The lines fitted with double peaked profiles are marked with ``D''.
(F) The lines fitted with flat-topped profiles are marked with ``F''.
(B) The blended transitions are marked with ``B''.
(S) Transitions that are smoothed to improve signal to noise ratio are marked with ``S''.  }

\end{landscape}

}


The spectrum is dominated by the strong transitions from five species: 
	SiS, HC$_3$N, HC$_5$N, HC$_7$N and c-C$_3$H$_2$. 
The peak intensities of SiS, HC$_3$N and HC$_5$N lines are larger than 120 mK. 
The peak intensities of HC$_7$N and c-C$_3$H$_2$ lines are larger than 50 mK. 
The peak intensities of other lines are about 20 mK. 
Weak lines with the peak intensities smaller than 20 mK have been smoothed 
	to have a channel width of 3.0 -- 3.6 $\ \mathrm{km\ s^{-1}}$ to improve signal to noise ratios. 
The smoothed lines are marked with red ``smoothed'' in the upper left of the corresponding panels in Figure \ref{Fig1a} and marked with ``S'' in Table \ref{Parameters}.
The rms noise of peak intensity, which is estimated 
	by fitting the baseline to a line or polynomial
	from frequency ranges around each particular transition 
	where the spectrum is apparently free of lines,  
	is about 3 -- 7 mK in 1.5 -- 3.6 $\mathrm{km\ s^{-1}}$ wide channel. 
Since the terminal expansion velocity of IRC +10216 have been estimated 
	in $\sim 14.5\ \mathrm{km\ s^{-1}}$ in previous studies.
The error of integrated intensity given in this work, 
	which is the product of the rms noise of peak intensity and 
	the width of the line for about $29\ \mathrm{km\ s^{-1}}$,
	is about 100 -- 200 $\mathrm{mK\ km\ s^{-1}}$.
The lines in this work reveal an average LSR velocity of about $-26.7\pm2.0\ \mathrm{km\ s^{-1}}$ 
	and an average terminal expansion velocity of about $13.9\pm2.0\ \mathrm{km\ s^{-1}}$, 
	which are consistent with previous studies
	(e.g., $-26.5\pm0.3\ \mathrm{km\ s^{-1}}$ and $14.5\pm0.2\ \mathrm{km\ s^{-1}}$, respectively \citep{cernicharo2000lambda}, $-26.404\pm0.004\ \mathrm{km\ s^{-1}}$ and $13.61\pm0.05\ \mathrm{km\ s^{-1}}$, respectively \citep{he2008spectral}). 

\section{Discussion}

  \begin{figure}
   \centering
    \includegraphics[height=\hsize]{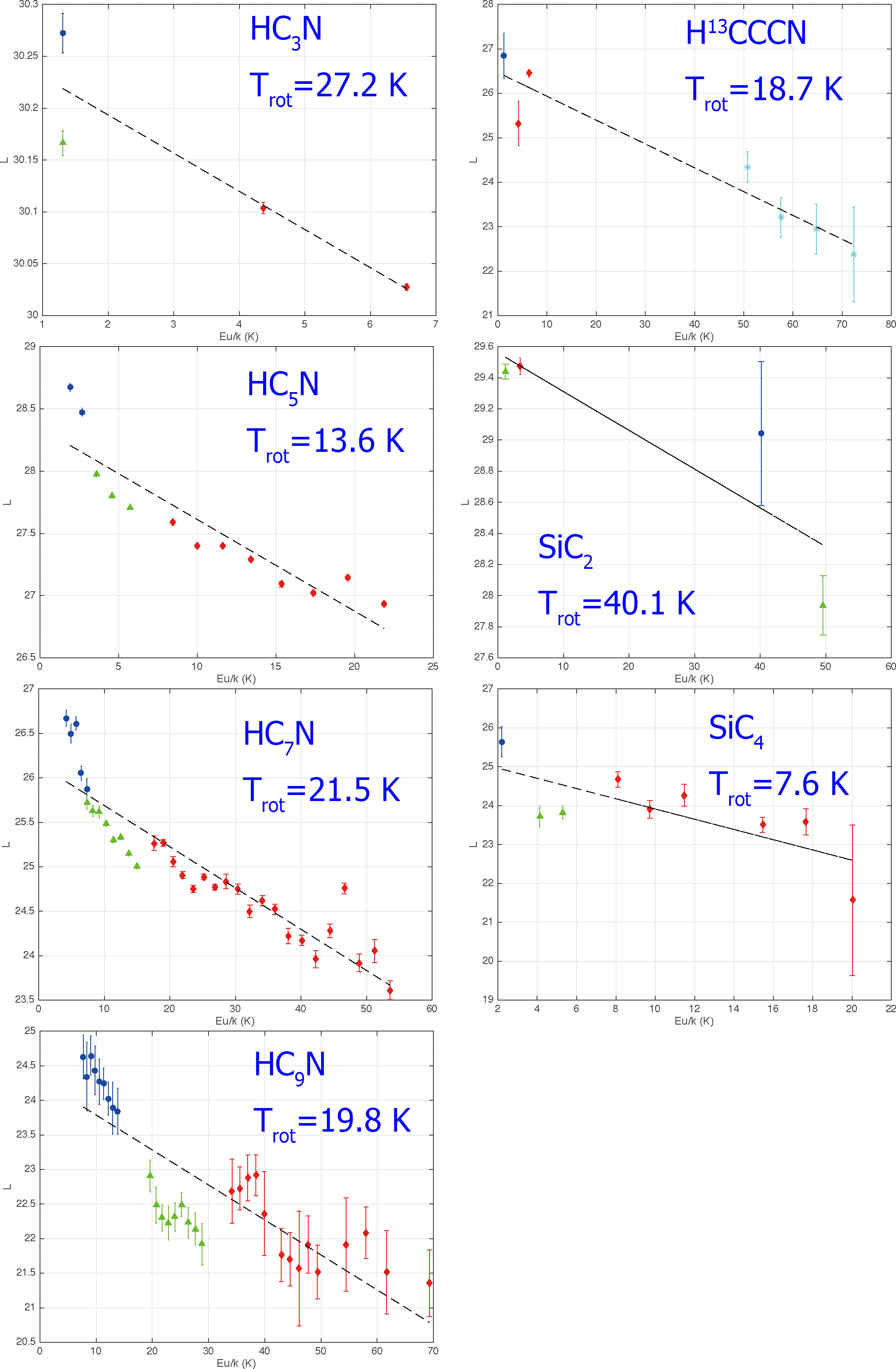}
        \caption{Rotational diagrams for the observed molecules in IRC +10216. 
   The variable L denotes the left-hand side of equation (\ref{E2}).
   Black dashed lines represent linear least-squares fit to the rotational diagram. 
   The blue circles are from our TMRT-65 m observations. 
   The green triangles are obtained from \citet{gong20151}. 
   The red diamonds are obtained from \citet{kawaguchi1995spectral}. 
   The magenta cross are obtained from \citet{remijan2007detection}. 
   The cyan asterisks are obtained from \citet{he2008spectral}. 
   Their values have been corrected for beam dilution. 
   The molecules and their corresponding rotational temperatures are given in each panel.
  }
    \label{Fig2}%
    \end{figure}
    \addtocounter{figure}{-1}
 \begin{figure}
   \centering
\includegraphics[height=\hsize]{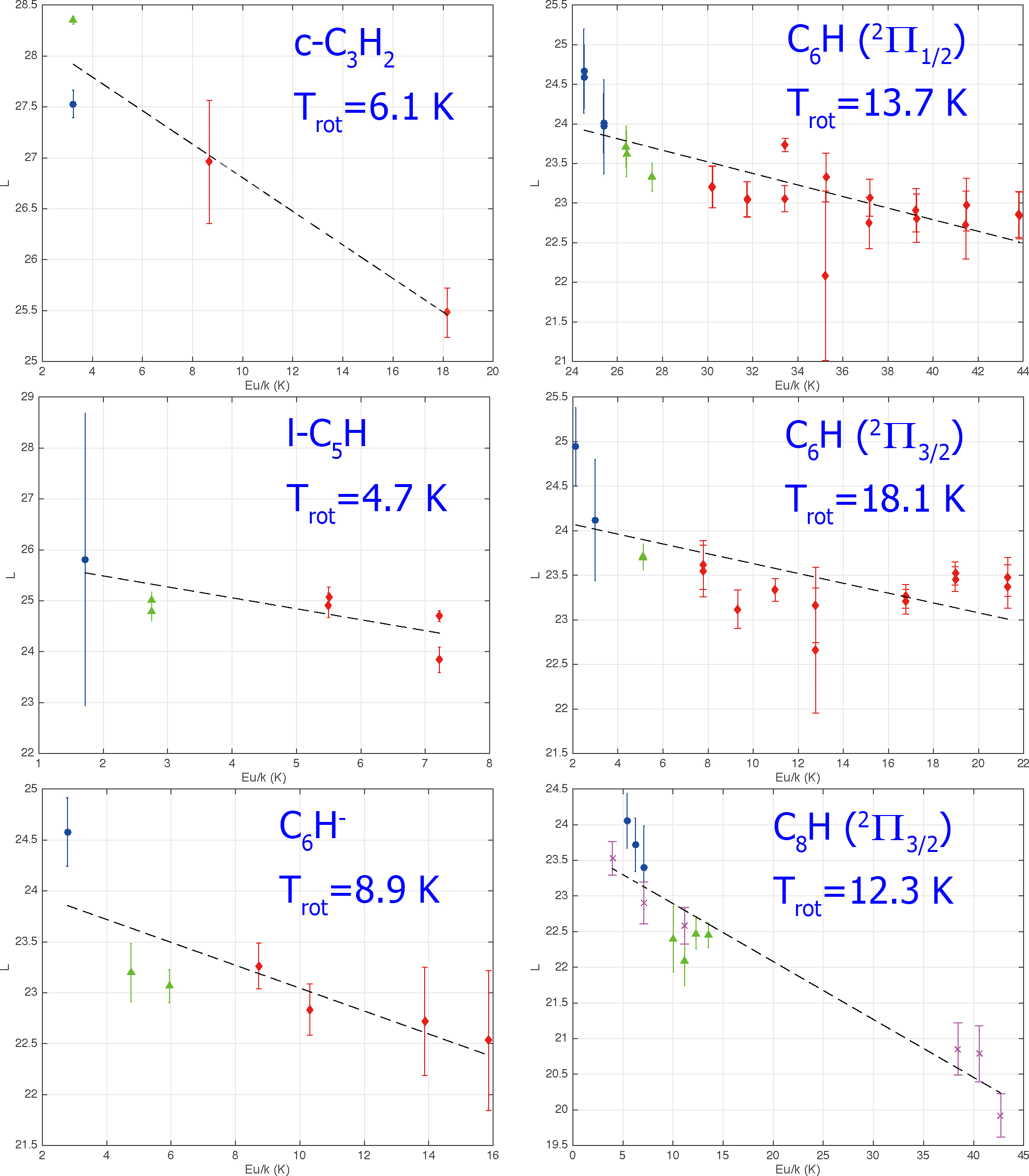}
\caption{continued.}
 \end{figure}

It was suggested that single-peaked lines arise from optically thick transitions, 
	while flat-topped and double-peaked lines arise from optically thin spatially unresolved and resolved transitions, 
	respectively \citep{Olofsson1982High,kahane1988carbon}. 
There are 11 double peaked lines and 19 flat-topped lines among the 30 unblended lines detected in this work. 
Therefore, we can assume optically thin for the lines detected in this work. 
By assuming LTE, rotational temperatures and column densities can be estimated from the rotational diagrams. 
The equation \citep{cummins1986survey}:
  \begin{equation} \label{E1}
      N = \frac{3kW}{8\pi^3\nu S\mu^2} Q(T_{rot}) \frac{T_{rot}}{T_{rot}-T_{bg}} \exp{(E_U/{k T_{rot} })} \,
   \end{equation}
gives the relation between the column density and the line intensity, 
	where $k$ is the Boltzmann constant, 
	$W\ (\int {T_{R}} \,\mathrm{d}\upsilon,\ \mathrm{K\ km\ s^{-1}})$ is the observed line integrated intensity, 
	$\nu\ (\mathrm{Hz})$ is the frequency of the transition, 
	$S\mu^2$ is the product of the total torsion-rotational line strength and the square of the electric dipole moment. 
$T_{rot}$ and $T_{bg}\ (2.73\ \mathrm{K})$ are the rotational temperature and background brightness temperature, respectively. 
$E_u$ is the upper level energy, and $Q(T_{rot})$ is the partition function. 
Values of $E_U/k$ and $S\mu^2$ are taken from the ``Splatalogue'' spectral line catalogs.

From equation (\ref{E1}), the formula for rotational diagrams is:
\begin{equation} \label{E2}
     \ln{\frac{3kW}{8\pi^3\nu S\mu^2}} = \ln{\frac{N}{Q(T^{*}_{rot})}}-\frac{E_U}{k T_{rot}} \,
   \end{equation}
   
where $Q(T^{*}_{rot})=Q(T_{rot})\frac{T_{rot}}{T_{rot}-T_{bg}}$.
To determine rotational temperatures with this method, 
 	there must be at least two transitions of the same molecule 
	with significant rotational temperature differences. 

SiS $(1-0)$ is a maser in IRC +10216 \citep{henkel1983sis} and 
 	its populations must deviate from LTE, 
	therefore this transition is excluded from the fitting. 
 The lines of HC$^{13}$CCN $(v=0,\ J=2-1)$ and HCC$^{13}$CN $(v=0,\ J=2-1)$ are blended 
 	owing to large uncertainties of intensities, so are also excluded. 
 Since only one line is detected in this work for each of 
 	HC$_3$N, H$^{13}$CCCN, c-C$_3$H$_2$, SiC$_2$, SiC$_4$ and C$_6$H$^-$, 
	these lines can not be used to determine rotational temperatures directly.
 For molecules of HC$_5$N, HC$_7$N, HC$_9$N, C$_6$H, C$_8$H and l-C$_5$H, 
 	although at least two lines are detected, 
	they do not have a wide dynamic range in upper level energies. 
 The 17.8 GHz to 26.3 GHz data from \citet{gong20151}, 
 	the 28 GHz to 50 GHz data from \citet{kawaguchi1995spectral}, 
	the C$_8$H data from \citet{remijan2007detection} and 
	the H$^{13}$CCCN data from \citet{he2008spectral}  
	are adapted as the complementary to data in the current survey 
	to derive rotational temperatures more precisely.

The integrated intensity ($\int {T_{MB}} \,\mathrm{d}\upsilon$, $\mathrm{mK\ km\ s^{-1}}$) of \citet{gong20151}
	 is obtained from the integrated flux density ($\int{S_{\nu}}\ \mathrm{d}\upsilon$, $\mathrm{mJy\ km\ s^{-1}}$), which is taken from Table 3 in \citet{gong20151}. 
And the conversion factor from the flux density ($S_{\nu}$, $\mathrm{Jy}$) to 
	the main beam brightness temperature ($T_{mb}$, $\mathrm{K}$) is $T_{mb}/S_{\nu}\sim 1.5\ \mathrm{K/Jy}$ at 22 GHz.
In \citet{gong20151}, across the whole frequency range, the beam size is $35\arcsec ?- 50\arcsec$ ($\sim 40\arcsec$ at 23 GHz).
The rms noise of intensity is estimated in the same way as in this work
	using online-data of the observed spectrum.
In \citet{kawaguchi1995spectral}, the main beam efficiency ($\eta$) was measured to be
	$0.78\pm0.06$ at 30 GHz, $0.79\pm0.05$ at 43 GHz and $0.71\pm0.07$ at 49 GHz.
The main beam efficiency at the other frequencies was assumed to be 0.78 in the region between 28 and 35 GHz,
	and was interpolated form the measured values in the region between 35 and 50 GHz.
The measured beam size (FWHM) was $34.9\arcsec \pm 1.1\arcsec$ at 49 GHz and $41.5\arcsec \pm 1.1\arcsec$ at 43 GHz.
The brightness temperature ($T_R$) of the molecular transition in the source 
	is related to the antenna temperature ($T_A^*$) as $T_R=T_A^*/(\eta\ \eta_{BD})$.
The integrated intensity ($\int{T_A^*}\ \mathrm{d}\upsilon$) and the rms noise of antenna temperature
	are taken from Table 1 in \citet{kawaguchi1995spectral}.
In \citet{remijan2007detection}, the beam size (FWHM) is approximated by $\theta_{beam}=740\arcsec/\nu(\mathrm{GHz})$.
The main beam efficiency ($\eta$), the integrated intensity ($\int{T_A^*}\ \mathrm{d}\upsilon$) and the error to the integrated intensity
	are taken from Table 1 in \citet{remijan2007detection}.
In \citet{he2008spectral}, the FWHM beam size of the KP12M is 43\arcsec at 145 GHz.
The integrated intensity ($\int{T_{MB}}\ \mathrm{d}\upsilon$) and the rms noise of main beam temperature
	are taken from Table 10 in \citet{he2008spectral}.
The errors to the integrated intensities of \citet{gong20151}, \citet{he2008spectral} and \citet{kawaguchi1995spectral} are computed in the same way with this work.


\begin{table*}
\caption{Column densities and rotational temperatures of the molecules in IRC +10216.}
\centering
\label{Densities}
\begin{tabular}{ccccc}
\hline\hline
Species & $T_{rot} (\mathrm{K})$ & $N\ \mathrm{cm^{-2}} $ & $X(N/N_{H_2})$ & Ref.\\
\hline
\multirow{3}{*}{HC$_3$N} & $27.2\pm19.4$ & $(1.94\pm0.21)\times10^{15}$ & $9.24\times10^{-7}$ & 0\\
 & $24.7\pm18.5$ & $(1.4\pm0.2)\times10^{15}$  & $6.7\times10^{-7}$ & 1\\
&26&$1.7\times10^{15}$ & &2\\
 \hline
H$^{13}$CCCN &$18.7\pm3.3$ & $ (3.25\pm1.46)\times10^{13}$ & $1.55\times10^{-8}$ &0 \\
\hline
 \multirow{3}{*}{HC$_5$N} &  $13.6\pm2.1$ &$(5.47\pm0.77)\times10^{14}$ & $2.61\times10^{-7}$ &0 \\
& $18.8\pm1.3$ & $(4.6\pm0.2)\times10^{14}$ & $2.2\times10^{-7}$ & 1\\
& $27\pm5$ & $(2.7\pm0.2)\times10^{14}$ & &2 \\
\hline
\multirow{3}{*}{HC$_7$N }& $21.5\pm1.7$&$(6.21\pm0.67)\times10^{14}$ &$2.96\times10^{-7}$ &0\\
 & $12.1\pm1.3$ & $(3.7\pm0.4)\times10^{14}$ &$1.8\times10^{-7}$ & 1 \\
 & $26\pm2$ & $(1.52\pm0.07)\times10^{14}$ & &2\\
 \hline
\multirow{3}{*}{HC$_9$N} & $19.8\pm2.4$ & $(5.88\pm1.28)\times10^{13}$  & $2.80\times10^{-8}$ &0 \\
 & $20.9\pm10.7$ & $(2.5\pm1.4)\times10^{13}$ & $1.2\times10^{-8}$ & 1 \\
 & $23\pm8$ & $(2.7\pm0.9)\times10^{13}$ & & 2\\
\hline
 \multirow{2}{*}{c-C$_3$H$_2$}& $6.1\pm2.8$ & $(1.39\pm1.07)\times10^{14}$ & $6.60\times10^{-8}$  & 0 \\
& $5.5\pm0.6$ &  $(2.1\pm0.4)\times10^{14}$ & $9.9\times10^{-8}$ & 1 \\
\hline
\multirow{3}{*}{l-C$_5$H $^2\Pi_{1/2}$}&$ 4.7\pm2.0$ & $(2.96\pm1.21)\times10^{14}$ &$1.41\times10^{-7}$& 0\\
& $8.3\pm2.0$ &$ (2.9\pm0.6)\times10^{13}$ & $1.4\times10^{-8} $& 1\\
& $27\pm5$  & $(2.9\pm0.3)\times10^{14}$& & 2\\
\hline
\multirow{2}{*}{C$_6$H$^-$} &$8.9\pm3.4$ &$(6.11\pm2.99)\times10^{12}$ &  $2.91\times10^{-9}$ &0 \\
& $26.9\pm4.0$ & $(5.8\pm0.5)\times10^{12}$ &  $2.8\times10^{-9}$ & 1\\
\hline
\multirow{3}{*}{C$_6$H $^2\Pi_{1/2}$} & $13.7\pm2.7$ & $(1.85\pm0.90)\times10^{14}$ &  $8.82\times10^{-8}$ &0 \\
& $20.7\pm2.4$ & $(1.0\pm0.2)\times10^{14}$ &$4.8\times10^{-8}$ & 1\\
& $46\pm4$  & $(1.13\pm0.6)\times10^{14}$ & & 2\\
\hline
\multirow{3}{*}{C$_6$H $^2\Pi_{3/2}$} &$18.1\pm5.8$ & $(5.38\pm1.18)\times10^{13}$ & $2.56\times10^{-8}$ &0 \\
& $47.2\pm10.3$ & $(1.0\pm0.1)\times10^{14}$ &  $4.8\times10^{-8}$ & 1 \\
& $35\pm2$  & $(1.65\pm0.07)\times10^{14}$ &  &2 \\
\hline
\multirow{2}{*}{C$_8$H $^2\Pi_{3/2}$}&$12.3\pm1.6$ & $(5.00\pm1.08)\times10^{13}$ & $2.38\times10^{-8}$ &0 \\
& $13.9\pm1.3$ & $(8.4\pm1.4)\times10^{12}$ & $4.0\times10^{-9}$ &1\\
\hline
\multirow{3}{*}{SiC$_2$}& $40.1\pm36.2$ & $(1.87\pm1.34)\times10^{15}$ &  $8.90\times10^{-7}$ & 0\\
 & $31.8\pm0.9 $ & $(1.2\pm0.0)\times10^{15}$ & $5.7\times10^{-7}$ &1\\
 & 16 & $2.2\times10^{14}$ & &2\\
\hline
\multirow{2}{*}{SiC$_4$}& $7.6\pm3.0$ & $(1.46\pm0.92)\times10^{13}$ &  $6.98\times10^{-9}$ & 0 \\
& $22.3\pm17.5$ &  $(1.1\pm0.4)\times10^{13}$ &  $5.2\times10^{-9}$  & 1\\
\hline
\end{tabular}
\tablefoot{
References for rotational temperatures and column densities from : (0) This work; (1) \citet{gong20151}; (2) \citet{kawaguchi1995spectral}.
The results of \citet{gong20151} and \citet{kawaguchi1995spectral} are also used the method of rotational diagrams. \\
}
\end{table*}

The intensities of the detected lines should be divided by $\theta^2_s/(\theta^2_s+\theta^2_{beam})$ for beam dilution 
	to derive the physical parameters, 
	where $\theta_s$ is the source size, and $\theta_{beam}$ is the beam size \citep{bell1993the}. 
So, the brightness temperature ($T_{R}$) 
	of the molecular transition in the source is related to the main beam temperature ($T_{MB}$) as 

\begin{equation} 
    T_{R}=T_{MB}/\eta_{BD}=T_{MB}\frac{\theta^2_{beam}+\theta^2_{s}}{\theta^2_{s}} \,
   \end{equation}

The source sizes are taken based on previous high resolution mapping of different molecules toward IRC +10216 by interferometer.
For species without high resolution mapping, their sizes are taken to be the same as chemically related species. 
Therefore, source sizes of HC$_3$N and HC$_5$N are $30\arcsec$ that determined by new JVLA observations. 
Source size of H$^{13}$CCCN is taken to be the same as that of HC$_3$N. 
Source sizes of HC$_7$N and HC$_9$N are taken to be the same as that of HC$_5$N. 
The sizes of l-C$_5$H, c-C$_3$H$_2$, C$_6$H, C$_6$H$^-$ and C$_8$H are taken to 
	be the same as that of C$_4$H \citep{guelin1993mgnc}, which is $30\arcsec$. 
The size of SiC$_2$ is $27\arcsec$, and the source size of SiC$_4$ is assumed to be the same as it \citep{lucas1995plateau}. 
The size of SiS is $18\arcsec$ by IRAM Plateau de Bure interferometer \citep{lucas1995plateau}. 
We carried out linear least-square fits to the rotational diagrams of 12 species. 
The results are shown in Figure \ref{Fig2} and Table \ref{Densities}. The results of \citet{gong20151} and \citet{kawaguchi1995spectral} in Table \ref{Densities} are obtained from the methods of rotational diagrams as well.

This work assumes the same source sizes as \citet{gong20151}, 
	and the H$_2$ average column density ($N_{H_2}$) of $2.1\times10^{21}\ \mathrm{cm^{-2}}$ within a typical radius of $15\arcsec$ taken from \citet{gong20151} is used to calculate molecular fractional abundances relative to H$_2$.
The derived rotational temperatures ($T_{rot}$), column densities ($N$) and 
	molecular fractional abundances relative to H$_2$ ($X(N/N_{H_2})$) together 
	with results from the literature, are listed in Table \ref{Densities}. 
The column densities of the molecules range from $10^{12}$ to $10^{15}\ \mathrm{cm^{-2}}$, 
	and the fractional abundances relative to H$_2$ of the species detected in Ku band range 
	from $2.91 \times 10^{-9}$ to $9.24\times10^{-7}$ in IRC +10216. 

The blue-shifted component of the SiS ($v=0,\ J=1-0$) line shown in Figure \ref{Fig1d}(d) 
	is stronger than the red-shifted one 
	due to maser amplification (18154.9 MHz; \citealt{henkel1983sis}). 
The peak intensity ratio of the blue-shifted component to the red-shifted component 
	is estimated to be $1.97\pm0.02$ in \citet{gong20151}. 
In this work, the peak intensity ratio is about $1.20\pm0.02$, 
	which is smaller than the ratio observed in \citet{gong20151}. 
The reason for the difference of the peak intensity ratio of SiS may be that 
	the velocity resolution at 18154.888 MHz of 1.5 $\mathrm{km\ s^{-1}}$ in the work is bigger than that of 1.008 $\mathrm{km\ s^{-1}}$ in Gong et al. (2015).
Lines of HC$_7$N ($v=0,\ J=16-15$), HC$_3$N ($v=0,\ J=2-1$) and c-C$_3$H$_2$ ($v=0,\ J=1(1,\ 0)-1(0,\ 1)$) have also been detected in \citet{gong20151}. 
 The parameters of HC$_7$N ($v=0,\ J=16-15$), SiS ($v=0,\ J=1-0$),  
	HC$_3$N ($v=0,\ J=2-1$) and c-C$_3$H$_2$ ($v=0,\ J=1(1,\ 0)-1(0,\ 1)$) in this work are shown in Table \ref{Parameters}.
The observed properties of these lines are displayed in Table \ref{ParaRatios}. 
	
The integrated intensities ($\int {T_{R}} \,\mathrm{d}\upsilon$) are shown in Table \ref{Ratios1}. 
The integrated intensities of HC$_7$N ($v=0,\ J=16-15$), SiS ($v=0,\ J=1-0$) 
	and HC$_3$N ($v=0,\ J=2-1$) in this work are consistent with 
	the results derived from \citet{gong20151}. 
But the integrated intensity of c-C$_3$H$_2$ ($v=0,\ J=1(1,\ 0)-1(0,\ 1)$) 
	in this work is much smaller than that in \citet{gong20151}. 
Since the c-C$_3$H$_2$ ($v=0,\ J=1(1,\ 0)-1(0,\ 1)$) line is at the edge of frequency range covered by the receiver, the flux calibration is less reliable there.

Light variability may also affect the accuracy of the integrated intensities.
Intensity comparisons are more difficult for weaker lines.
And for abundant species and their isotopologues 
	toward the archetypical circumstellar envelope of IRC+10216, 
	based on the existed observations \citep{cernicharo2000lambda,Cernicharo2014Discovery},
	some line intensities of the high rotational lines may follow the infrared flux variations, 
	and some line intensities of the low-J transitions may not.
But there is no direct evidence indicating 
	line intensities of the emission lines detected 
	in this work, \citet{gong20151} and \citet{kawaguchi1995spectral} 
	correlate with the continuum intensity. 
Since most of the detected lines in this work are weak, 
	and the data for the strong transitions in this work 
	is not enough to study the effect of the variability on the integrated intensity. 
Therefore, it is assumed that the periodic variations of the stellar IR flux do not modulate molecular line emission here.


\begin{table*}
\centering
\caption{Line Parameters of HC$_7$N ($v=0,\ J=16-15$), SiS ($v=0,\ J=1-0$), HC$_3$N ($v=0,\ J=2-1$) and c-C$_3$H$_2$ ($v=0,\ J=1(1,\ 0)-1(0,\ 1)$) detected by This work and \citet{gong20151}.}             
\label{ParaRatios}    
\begin{tabular}{cccc cccc}       
\hline\hline             
\multirow{2}{*}{Species}&\multirow{2}{*}{Transitions}& \multirow{2}{*}{Rest Freq. (MHz)}&\multicolumn{2}{c}{$\int {T_{MB}} \,\mathrm{d}\upsilon\ (\mathrm{mK\ km\ s^{-1}})$} & \multirow{2}{*}{$\theta_s$ \ ($\arcsec$)} & \multicolumn{2}{c}{$\theta_{beam}$ \ ($\arcsec$)}  \\ 
\cline{4-5} 
\cline{7-8} 
& & & This Work & Gong &  & This Work & Gong\\
\hline                        
HC$_7$N &$v=0,\ J=16-15$ &18047.970& 1629(194) & 1577(116)& 30 & 54 & 50 \\
SiS & $v=0,\ J=1-0$ &18154.888& 6692(186) & 7140(110)& 18 & 53  & 49  \\    
HC$_3$N &$v=0,\ J=2-1$ &18196.226& 10337(197) & 10445(125) & 30 & 53 & 49  \\
c-C$_3$H$_2$&$v=0,\ J=1(1,\ 0)-1(0,\ 1)$ &18343.143 & 1161(157) & 2973(116)& 30 & 53 & 49  \\ 
\hline                              
\end{tabular}
\end{table*}


\begin{table*}
\caption{The integrated intensity ratios (This Work: \citet{gong20151}) of HC$_7$N ($v=0,\ J=16-15$), SiS ($v=0,\ J=1-0$), HC$_3$N ($v=0,\ J=2-1$) and c-C$_3$H$_2$ ($v=0,\ J=1(1,\ 0)-1(0,\ 1)$) }             
\label{Ratios1}      
\centering                          
\begin{tabular}{c c c  c c}        
\hline\hline                 
\multirow{2}{*}{Species}& \multirow{2}{*}{Rest Freq. (MHz)}& \multicolumn{2}{c}{$\int {T_{R}} \,\mathrm{d}\upsilon\ (\mathrm{mK\ km\ s^{-1}})$} & Ratios \\   
\cline{3-4}
&  &This Work &\citet{gong20151}& This: Gong\\
\hline                        
HC$_7$N & 18047.9697& 6907(823) & 5958(438)&  $1:(0.863\pm0.121)$\\
SiS& 18154.8880& 64710(1799) & 60051(925)& $1:(0.928\pm0.029)$ \\     
HC$_3$N &18196.2260& 42600(812) & 38309(458) &$1:(0.899\pm0.020)$ \\
c-C$_3$H$_2$& 18343.1430& 4784(647) & 10904(425)&$1:(2.279\pm0.321)$ \\
\hline                                   
\end{tabular}
\end{table*}

  \begin{figure}   
   \centering
\resizebox{\hsize}{!}{\includegraphics{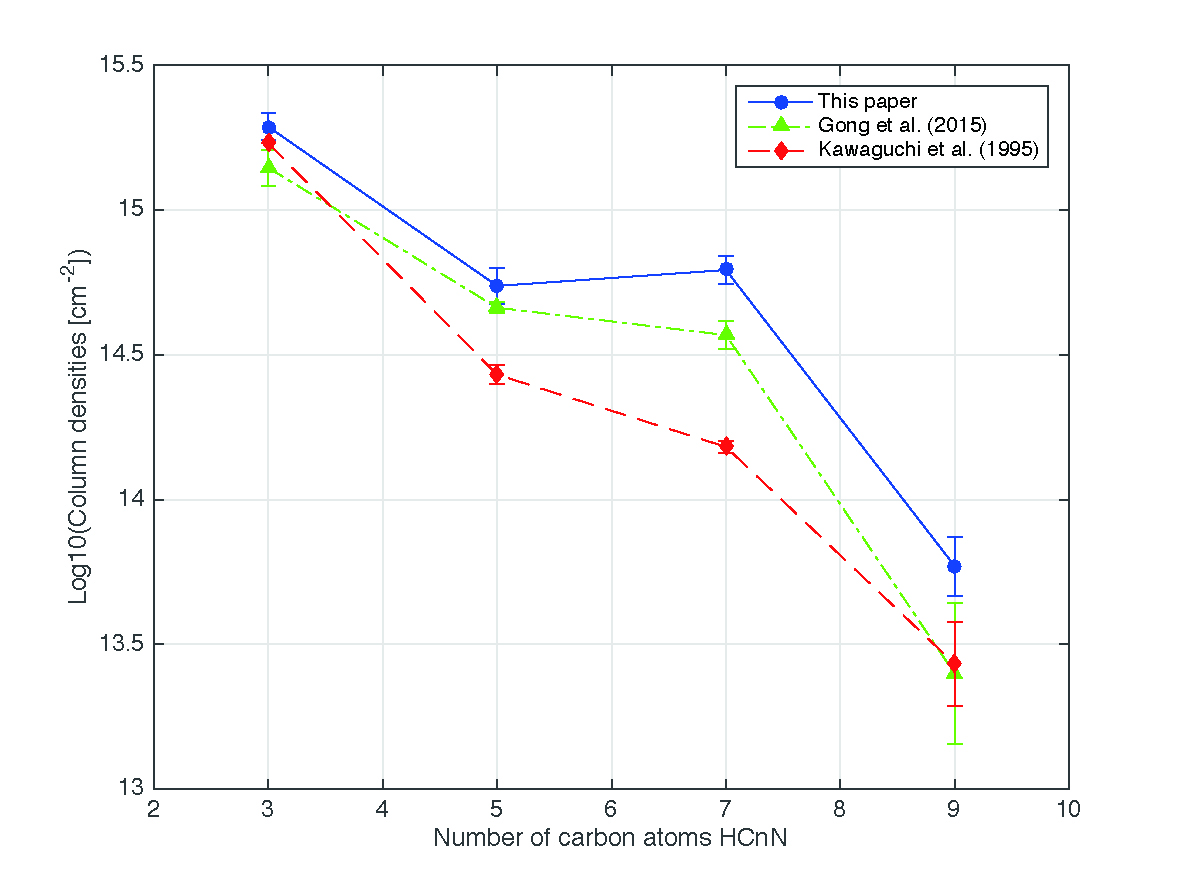}} 
   \caption{Comparison of the column densities. 
   The circles with error bars are from our TMRT-65 m observations. 
   The triangles with error bars are obtained from \citet{gong20151}.
   The asterisks without error bars are obtained from \citet{kawaguchi1995spectral}.  
   }
    \label{Fig3}%
    \end{figure}

  \begin{figure}  
   \centering
    \includegraphics{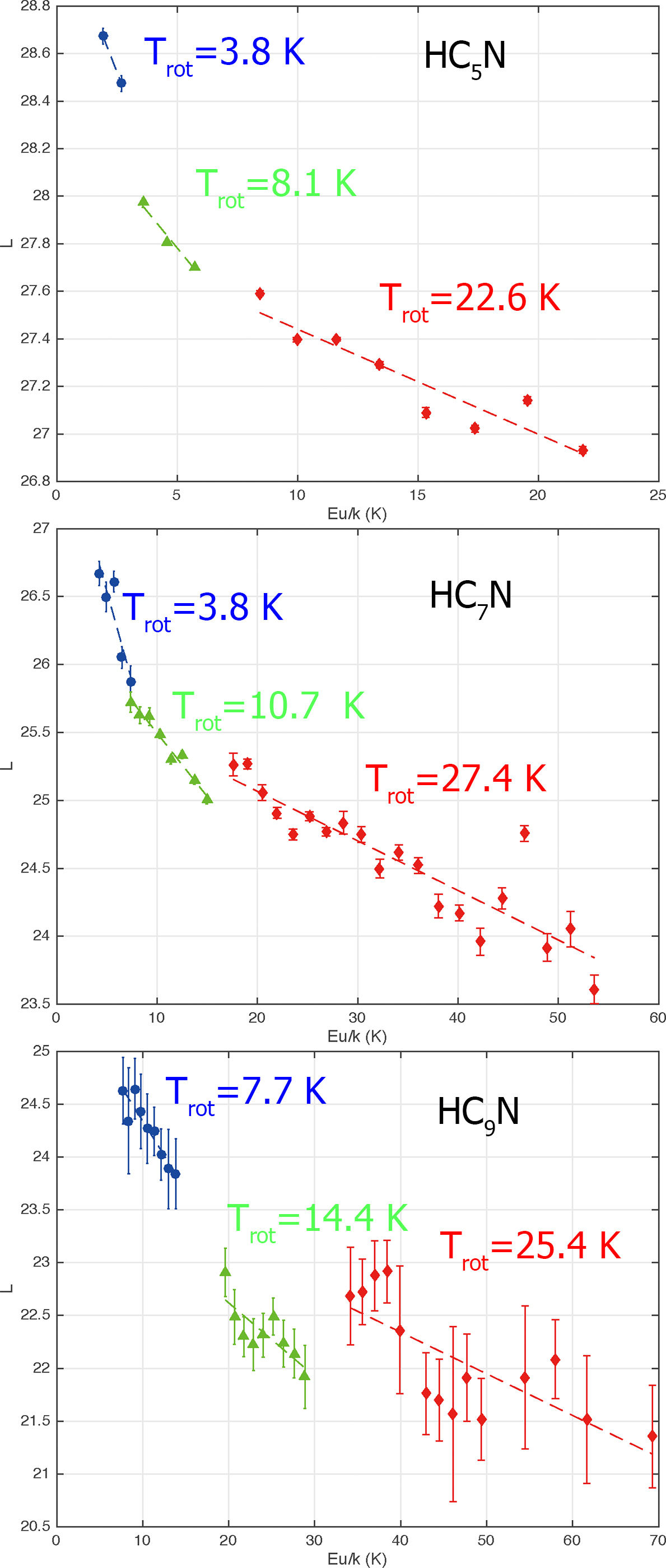}  
   \caption{The red circles are from our TMRT-65 m observations. 
   The green triangles are obtained from \citet{gong20151}. 
   The blue asterisks are obtained from \citet{kawaguchi1995spectral}. 
   The dashed lines of different colors represent linear least-squares fit 
   	to the rotational diagram accounting for data obtained from corresponding surveys. 
   }
    \label{Fig4}%
   \end{figure}

   \begin{table*}
\caption{The abundance ratios ( HC$_3$N: HC$_5$N: HC$_7$N: HC$_9$N ) of this work, \citet{gong20151} and \citet{kawaguchi1995spectral}}             
\label{Ratios2}      
\centering                          
\begin{tabular}{c c c c c}        
\hline\hline                 
Species & HC$_3$N &HC$_5$N  &  HC$_7$N & HC$_9$N  \\    
\hline                        
 This work&  100 & (28.2$\pm$5.0) & (32.0$\pm$4.9) & (3.0$\pm$0.8) \\      
 \citet{gong20151} &100 & (32.9$\pm$4.9)  &(26.4$\pm$4.7)  & (1.8$\pm$1.0) \\
 \citet{kawaguchi1995spectral}  &100 & (15.9$\pm$1.2) & (8.9$\pm$0.4)    & (1.6$\pm$0.5) \\
\hline                                   
\end{tabular}
\end{table*}

The source sizes of HC$_{2n+1}$N assumed in \citet{kawaguchi1995spectral} are 
	the same with those in this paper and \citet{gong20151}.
The column densities of HC$_{2n+1}$N among this paper, 
	\citet{gong20151} and \citet{kawaguchi1995spectral} 
	in Table \ref{Densities} can be compared directly,
	shown in Figure \ref{Fig3}. 
The column densities of HC$_3$N, HC$_5$N , HC$_7$N and HC$_9$N in this work are greater 
	than the densities derived from \citet{gong20151} and \citet{kawaguchi1995spectral}.  	
The abundance ratios for HC$_3$N: HC$_5$N: HC$_7$N: HC$_9$N are shown in Table \ref{Ratios2}. 
The ratio in this work is consistent with the ratio derived from \citet{gong20151}, 
	but is greater than the ratio in \citet{kawaguchi1995spectral}. 
The abundance ratio for C$_6$H $^2\Pi_{3/2}$: C$_8$H $^2\Pi_{3/2}$ is calculated to be $(1.1\pm0.3): 1$, 
	which is much smaller than the ratio observed in \citet{gong20151}, $(11.9\pm2.3): 1$. 
The obvious difference in abundances may suffer from low data quality in lines of C$_6$H and C$_8$H.
	
The transitions of HC$_3$N and its $^{13}$C substitutions are optically thin. 
Thus, the isotopic ratio can be directly obtained from their integrated intensity ratio.
The derived $^{12}$C/$^{13}$C ratios are $32\pm16$ from [HCCCN]/[H$^{13}$CCCN] ($v=0,\ J=2-1$). 
This is much smaller than 71 obtained from [$^{12}$CO]/[$^{13}$CO] by \citet{ramstedt2014the} and 
	is also smaller than $49\pm9$ derived from HC$_5$N and its $^{13}$C isotopologues by \citet{gong20151}. 
This value agrees with $34.7\pm4.3$ derived from [SiCC]/[Si$^{13}$CC] by \citet{he2008spectral}.
Since, there is only one transition used for calculating the isotopic ratio in this work, 
more transitions need to be adapted as the complementary to derive the isotopic ratio more precisely.
  
\citet{bell1992sensitive} have reported that the HC$_5$N molecule 
	in the circumstellar envelope of IRC +10216 traces two molecular regions: 
	a warm one ($T_{ex} \sim 25\ \mathrm{K}$) traced by high-J transitions, 
	i.e., those with upper-state energies $E_U/k$ greater than 20 K, 
	and a cold region with $T_{ex} \sim 13\ \mathrm{K}$ traced by low-J transitions with $E_U/k < 10\ \mathrm{K}$. 
In Figure \ref{Fig4}, the rotational temperatures of HC$_5$N are estimated to be 
	This: Gong: Kawaguchi = 1: 2.1: 5.9.
The fitted rotational temperature of HC$_5$N for high-J transitions is greater. 
And the rotational temperatures ($T_{rot}$) of HC$_7$N and HC$_9$N are estimated to be 
	This: Gong: Kawaguchi = 1: 2.8: 7.2 and 1: 1.9: 3.3.
Similar as in the HC$_5$N case, the rotational temperatures of HC$_7$N and HC$_9$N for high-J transitions are also greater. 
Therefore, the high-J transitions of the HC$_5$N, HC$_7$N and HC$_9$N molecules in the circumstellar envelope of IRC +10216 trace the warmer molecular regions.
The abundance ratios of HC$_5$N, HC$_7$N and HC$_9$N derived from surveys of different frequency ranges are estimated to be 
	This: Gong: Kawaguchi = \text{100: 43: 55, 100: 24: 28 and 100: 30: 44}, respectively. 
Obvious differences exist in the abundance ratios derived from different transitions, and this result may suffer from opacity effects. 
It also suggests that the assumption of thermal equilibrium is not accurate to estimate rotational temperatures and column densities.
 %

\section{Summary}
A spectral line survey of IRC +10216 between 13.3 and 18.5 GHz 
      	is carried out using the TMRT. 
Forty-one spectral lines of 12 different molecules and radicals are detected in total. 
Several carbon-chain molecules are detected, 
	including HC$_3$N, HC$_5$N, HC$_7$N, HC$_9$N, C$_6$H, C$_8$H, 
	C$_6$H$^-$, l-C$_5$H, SiC$_2$, SiC$_4$ and c-C$_3$H$_2$.

The rotational temperatures and column densities of the detected molecules are derived by assuming LTE. 
Their rotational temperatures range from 4.7 to 40.1 K, 
      	and molecular column densities range from $10^{12}$ to $10^{15} \ \mathrm{cm^{2}}$. 
Molecular abundances relative to H$_2$ range between $2.91 \times 10^{-9}$ and $9.24\times10^{-7}$. 
From the comparison with previous works, 
	it is clear that the higher-J transitions of 
	the HC$_5$N, HC$_7$N and HC$_9$N molecules in the circumstellar envelope of IRC +10216 
	traces the warmer molecular regions.
And there are obvious differences in the abundance ratios derived from different transitions.

\begin{acknowledgements}
    This work is supported by the Natural Science Foundation of China (No. 11421303, 11590782) .
 \end{acknowledgements}
%

%
%

\end{document}